\documentclass[reprint,
 amsmath,amssymb,
 aps,
pra,
]{revtex4-1}

\usepackage[version=3]{mhchem}

\usepackage[colorinlistoftodos]{todonotes}

\usepackage{epstopdf}

\usepackage{graphicx}
\usepackage{dcolumn}
\usepackage{bm}

\usepackage{algpseudocode}
\usepackage{algorithm}

\usepackage{ulem}

\usepackage{mathtools}

\newcommand{\la}{\langle}
\newcommand{\ra}{\rangle}

\begin{document}

\normalem

\preprint{APS/123-QED}

\title{Quantum Control Landscapes Beyond the Dipole Approximation}


\author{Benjamin Russell}
\affiliation{Frick Laboritory, Princeton University, Princeton, NJ 08540}
\email{br6@princeton.edu}

\author{Rebing Wu}
\affiliation{Department of Automation, Tsinghua University, Beijing, China, 100084}
\email{rbwu@tsinghua.edu.cn}

\author{Herschel Rabitz}
\affiliation{Frick Laboritory, Princeton University, Princeton, NJ 08540}
\email{hrabitz@princeton.edu}

\date{\today}

\begin{abstract}
We investigate the control landscapes of closed, finite level quantum systems beyond the dipole approximation by including a polarizability term in the Hamiltonian.
Theoretical analysis is presented for the $n$ level case and formulas for singular controls, which are candidates for landscape traps, are compared to their analogues in the dipole approximation.
A numerical analysis of the existence of traps in control landscapes beyond the dipole approximation is made in the four level case.
A numerical exploration of these control landscapes is achieved by generating many random Hamiltonians which include a term quadratic in a single control field.
The landscapes of such systems are found numerically to be trap free in general.
This extends a great body of recent work on typical landscapes of quantum systems where the dipole approximation is made.
We further investigate the relationship between the magnitude of the polarizability and the magnitude of the controls resulting from optimization.
It is shown numerically that including a polarizability term in an otherwise uncontrollable system removes traps from the landscapes of a specific family of systems by restoring controllability.
We numerically assess the effect of a random polarizability term on the know example of a three level system with a second order trap in its control landscape.
It is found that the addition of polarizability removes the trap from the landscape.
The implications for laboratory control are discussed. 

\end{abstract}

\pacs{Valid PACS appear here} 
\maketitle

\section{Introduction}

There has been much recent interest in quantum control and quantum control landscapes \cite{Raj, in2, in3, atql, coarql, sscl, num2, num1, QCOCU, OCQSL, Wu, eql, ecql} arising from the desire to control quantum systems for novel technological applications.
This has been driven by a large amount of experimental and theoretical progress, specifically including the development of femtosecond timescale shaping of laser pulses in the laboratory.
Key experimental areas include the control of atomic systems and molecular systems \cite{fsmc}, the control of chemical reactions and bond manipulation \cite{hrbb} up to the scale of biological systems \cite{Her}.
One area of application which has attracted interest is quantum information processing \cite{QIP1, QIP2, QIP3} wherein optimal control can be used to design pulses to implement quantum gates with high fidelity and to minimize errors introduced by decoherence and environmental noise.
Typical desiderata in quantum control include, driving an initial density matrix $\rho$ to a desired density matrix, maximizing the expectation of a given observable $\la O \ra_{\rho}$ and driving the unitary propagator $U_t$ to a desired goal `gate' (in quantum information science) or, time evolution $W$.
In quantum information processing applications, one often seeks to minimize the time (especially in quantum information processing to avoid decoherence) to achieve a goal physical transformation while maximizing the fidelity of the transformation. 
We focus on the fidelity landscape (with a fixed end time significantly larger than the minimal time) rather than seeking results about the minimal time regime.
In some cases of quantum time optimal control, the associated control landscapes are known to have traps \cite{bon, TOsing} resulting from singular controls.
In this work we study the landscapes for the control of the quantum propagator of closed three and four level quantum systems with a single control field, extending existing studies by moving beyond the typical dipole approximation by including a polarizability term \cite{pols1,pols2} in the Hamiltonian.
This is motivated by the fact that this term is present in many physically realistic models of the control of quantum systems, including the control of complex molecules within which a control field can induce a redistribution of charge.
We specifically assess the potential for singular controls to introduce traps into the landscapes of such systems in order to understand under which conditions gradient based methodologies will succeed in discovering high fidelity controls.
This is motivated by the need to analyses which algorithms are appropriate for use in simulations and in automatic pulse discovery in the laboratory, and furthermore, to understand the fundamental nature of the topology of control landscapes of quantum systems in their own right.

Given a prescribed final time $T\in \mathbb{R}$ and a desired time evolution, or \emph{goal} gate, $W \in SU(n)$ (the special unitary Lie group) we measure the \emph{fidelity} of a time evolution as:
\begin{align}
J[U_T] = \frac{1}{N}\left|Tr(W^{\dagger}U_T) \right|^2
\label{tfid}
\end{align}
where $N=n^2-1$ (dimension of $\mathfrak{su}(n)$), so that the maximum of $J$ over all $U \in SU(n)$ is $1$.
We note that this is only one of several popular choices which also include $\widetilde{J}(U_T)=\Re \left(Tr(W^{\dagger}U_T)\right)$.
The critical point topology of the function $J$ is discussed in detail in \cite{jlan}, where it is shown that it possess only global maxima, global minima and saddle points.
The function $J$ possess no local optima as a function of $U\in SU(n)$.
We study the control landscape of the cost function: $F[E] = J[V_T[E]]$ where $E$ is the control field, and $V_T$ is the end-point map (see \cite{bon} for a more detailed and general discussion of this map in control theory).
$V_T$ is a mapping from the space of controls to the corresponding solution $U_T$ to the time dependent Schr\"odinger equation:
\begin{align}
\frac{d U_t}{dt} = -i H_t U_t
\label{sch}
\end{align}
Throughout this work, $\hbar$ has been set to $1$ unless indicated otherwise.
The type of Hamiltonian we study is as follows:
\begin{align}
H_t = H_0 + E(t) H_1 + E(t)^2 H_2
\label{polham}
\end{align}
where $iH_0, iH_1, i H_2 \in \mathfrak{su}(n)$.
This is the first step towards including higher order terms beyond the dipole approximation (where only the first power of $E$ is included) from the expansion:
\begin{align}
H_t =  H_0 + \sum_{k=1}^{\infty} E(t)^k H_k
\label{expa}
\end{align}
wherein $\{H_k\}$ asymptotically decay appropriately to ensure convergence.
The terms $H_k$ have a clear physical interpretation.
Specifically, the first additional term $H_2$ is the polarizability.
This term represents the ability of an external electromagnetic field to redistribute the charge with in a system so that a dipole is created.
In a more physically complete model of a molecular system interacting with an external field, the terms $H_k$ with scalar coefficients would be insufficient as higher order coupling will always be present to some extent.
The term $H_k$ would be replaced by an order $k+1$ tensor.
This physically motivates the investigation into the control landscapes of such systems.
Some work on the control of such systems can be found in \cite{Gri,rabpol,Tur,Tur2}.
For a physical discussion of this type of system and the interpretation of $H_3$ (`hyper polarizability') and the terms beyond this see \cite{hpo}.
While the control landscapes of quantum systems have been studied intensively, landscape analysis of systems which include this additional polarization term has not yet been performed.
For some spherically symmetrically systems, for example a hydrogen atom, the dipole $H_1$ is far smaller than the polarizability $H_2$.
In such cases, examining the role of polarizability as a tool for quantum control is a particularly salient direction of extension of existing work.

For a first example, a more fully physical description of a molecule in an external field would have to include the three components of that field $E_x, E_y, E_z$, each with their own couplings $H_{1,x}, H_{1,y},H_{1,z}$.
To assume the standard form of the Hamiltonian $H_0 + E(t)H_1$ is not only to assume that polarizability and all higher order couplings are negligible, but also to assume that the dipole of the molecule remains aligned with the external EM field throughout an evolution.

In this work we address the standard assumptions of quantum control landscape analysis (\ref{ass}) applied to systems which have a polarizability term present.
Firstly we assess many random quantum systems, with a polarizability term, for traps in their control landscapes.
It is seen that generically, no traps are present for initial controls near to the zero field.
We assess the affect of the addition of a polarizability on controllability of systems which would not otherwise be controllable.
Specifically, we see that it is highly typical for the addition of a random polarizability term to restore controllability and to thus remove traps from the landscape of a large family of such systems in the four level case.
We further analyse the effect of many different random polarizability terms on a known second order trap in the landscape of a specific three level system.
We find that all such terms cause the trap to dissapear.

\section{The Three Assumptions of Landscape Analysis}

\label{ass}

There are three assumptions underlying the analysis of quantum control landscapes.
These are not quite mathematical axioms, but they are more at the level of well established assumptions.
These assumptions are:
\begin{enumerate}
\item The system is controllable, i.e. every unitary $U_T$ can be implemented by some control $E$. Equivalently, the end-point map $V_T$ is globally surjective.
\item The Hesssian of $F$ is not negative definitive for any singular critical control. See (\ref{sings}) for definitions.
\item The controls are unconstrained, all control functions can be implemented without restriction on resources.
\label{axi}
\end{enumerate}
Based on several numerical and laboratory studies \cite{num1, num2} and mathematical analysis \cite{an1} these assumptions have been demonstrated to be sufficient conditions in practice to ensure that a quantum control landscape is trap free.
However, the weakest sufficient conditions for a trap free landscape are not known.
Assumption (3) in it's original form read: the derivative of the end point map is full rank everywhere on the landscape, however has been shown to be violated for some specific systems \cite{sscl} and potential effects on gradient based searches for optimal controls has been discussed in \cite{sqcs, Wu}.
Future work will include more exactly revising and refining assumption (3) in order to obtain necessary and sufficient conditions for a trap free landscape based on the geometric notion of \emph{transversality} of the end point map to the level sets of fidelity rather than local surjectivity of this map \cite{meunpub}.

In the context of quantum control (for systems without the polarizability term), it has been shown \cite{alt} that assumption (1) generically holds when $H_0, H_1$ are chosen uniformly at random.
It is shown that controllability (technically accessibility, i.e. every point on $SU(n)$ can be reached using some control at \emph{some} final time $T$) only fails for a null set of pairs $(iH_0, iH_1) \in \mathfrak{su}(n)\times \mathfrak{su}(n)$.
A controllability analysis of systems which include a polarizability term has been performed in \cite{Tur2} and controllability has been found to be similarly generic.

It has further been shown \cite{sscl} that there exists a critical time $T^*$ such that $\forall T \geq T^*$ the system is \emph{fixed-time controllable} as long as it is controllable.
That is to say that if $iH_0, iH_1$ generate $\mathfrak{su}(n)$ and the final time $T$ is large enough, one can always find a control $E$ such that $U_T = G$ for any goal operator in $SU(n)$.
As such, in all simulations a sufficiently large time has been chosen so that all systems are not restricted in accessibility by this factor.

As a result of these two observations, it is guaranteed that the first assumption will be satisfied for systems with Hamiltonians generated at random and controlled over sufficiently large intervals as the set for which this fails is null and thus the probability of generating such a case is $0$.
This is not to say that there are no uncontrollable systems in reality or that they cannot be deliberately mathematically constructed and studied, nor is it to say that they do not have interesting control landscape structure \cite{ecql} which can include traps.

The second assumption is the least well understood and has not been rigorously shown to hold for typical quantum systems.
It has been shown that in two level systems, singular controls never represent traps \cite{tf1, tf2} for the control of the density operator.
It is further known that, in randomly generated four level systems, singular controls are generically saddles within control space for the task of controlling the density matrix \cite{Wu}.
It has not however, despite mounting corroborating numerical evidence, been rigorously proven that the property that singular controls are always saddles in control space, rather than traps, always holds.

The third assumption is satisfied if no restriction is imposed on the control during simulations.
However, in laboratory practice, there are always restrictions on the control.
These restrictions are not only on the total achievable field fluence using any given device, but also on the ability to accurately implement and vary the control field.
The typical scenarios are those of the control of spins and molecules by electromagnetic fields.
These two control modes both have different signal to noise ratios which represent one practical type of restriction on the control which is not considered in this work.
Including noise in simulations is not simply a restriction on the space of controls considered, but is further a deviation from the assumed for of the underlying Schr\"odinger equation. To assess the effect of polarizabilty on robustness, a stochastic term would need to be included in the control field and we do not consider this scenario in this work.

As the first and third assumptions have been shown to hold generically, it is logical to investigate when assumption two can fail in order to investigate the effect of polarizability on landscape traps.

\section{Singular Controls}
\label{sings}

It was first discussed in \cite{Raj} that singular controls could in principle introduce traps into quantum control landscapes, but this was conjectured to be rare in practice.
This has since been backed up with extensive numerical analysis \cite{Wu,riv}.
Several studies \cite{atql, cat, sscl, riv, Wu} have discussed the potential effect of the existence of singular controls on quantum control landscapes and the significance of some of these findings has been debated \cite{coarql}.

A formula is known for the first variation of the end-point map with respect to the Hamiltonian.
If the time-dependent Hamiltonian $H_t$ of a (finite level) quantum system undergoes an arbitrary infinitesimal transformation $H_t \mapsto H_t + \delta H_t$ then the end-point map $V_T[iH_t] := U_T$ varies according to:
\begin{align}
U_T^{\dagger} \delta U_T = i \int_0^{T} U_t \delta H_t U_t^{\dagger} dt \in \mathfrak{su}(n)
\label{genvar}
\end{align}
In the case of the dipole approximation and a single control field, as is generically studied in quantum control landscape theory, this variation takes a specific form.
In the dipole approximation with a single control field, the Hamiltonian takes the form: $H_t = H_0 + E(t)H_1$ and thus a variation $\delta E$ of the control induces a corresponding variation $\delta H_t = \delta E(t) H_1$. By applying formula (\ref{genvar}) this yields:
\begin{align}
\label{abc2}
U_T^{\dagger}\delta U_T = \int_{0}^{T} \delta E(t) U_t^{\dagger} i H_1 U_t dt
\end{align}
A control $E$ is said to be \emph{singular} if there exists at least $B \in \mathfrak{su}(n)$ such that for all $\delta w$:
\begin{align}
\left\la U_T^{\dagger}\delta U_T, B \right\ra = 0
\label{singular}
\end{align}
where $\la \cdot, \cdot \ra$ is the trace (or any other real valued) inner product on $\mathfrak{su}(n)$.
Differentiating this expression twice with respect to $t$ yields an implicit formula for singular $E$ mathematically connecting such singular $E$ and singular trajectories $U_t$ in the case of a system in the dipole approximation \cite{Wu}.
Unfortunately, an explicit formula for the singular controls is not known and appears impossible to obtain by any method known to the authors or appearing in the literature.
Intuitively, a singular control is a control for which the end point cannot be `steered' in at least one particular direction on $SU(n)$ by applying a small variation to the control field.
It is noteworthy that, although a specific inner product is invoked here, the singularity of any given control does not depend on which inner product is chosen and any choice yields the same set of singular controls.
Singularity of a given $E$ is equivalent to the statement that the Fr\'echet derivative: $\delta V_T$ is rank deficient at the point $E$ in control space.
By substituting (\ref{genvar}) into (\ref{abc2}) and applying the the fundamental lemma of calculus of variations, one sees that a singular control must satisfy:
\begin{align}
\la U_t^{\dagger} i H_1 U_t, B \ra = 0, \ \ \ \forall t \in [0,T]
\end{align}
In the case of formula (\ref{polham}), where the Hamiltonian contains the additional polarizability term $E(t)^2 H_2$, the singular controls take a novel form.
Formula (\ref{singular}) in this case implies:
\begin{align}
\left\la \int_{0}^{T} \delta E(t) U_t^{\dagger} \left(i H_1 + 2E(t) i H_2 \right) U_t dt, B \right\ra = 0
\end{align}
which implies, again after applying the fundamental lemma of calculus of variations \cite{jost} and rearranging (assuming $\la U_t^{\dagger} i H_2 U_t, B\ra \neq 0$ $\forall t \in [0, T]$):
\begin{align}
E(t) = -\frac{1}{2}\frac{\la U_t^{\dagger} i H_1 U_t, B\ra}{\la U_t^{\dagger} iH_2 U_t, B\ra}
\label{sc}
\end{align}
which is in contrast to the form of the singular controls found in in \cite{Wu} where differentiation of formula (\ref{singular}) was required to determine the corresponding form of the singular controls.
This formula is applicable to the scenario of controlling the density matrix, rather than the propagator.
However, a similar formula can be found in the case of controlling the propagator and it requires an identical differentiation procedure to determine it.
At points in time where $\la U_t iH_2 U_t^{\dagger}, B\ra = 0$, another formula for $E(t)$ is needed.
This can be found by differentiation, analogous to the procedure found in \cite{bon} in a general form and applied to quantum control specifically in \cite{Wu}.
The number of derivatives required to find a singular control is known as the \emph{order} of a singular control and the quantity: $\text{dim}(SU(n)) - \text{rank}(\delta V_T)$ is known as the co-rank of a singular control.
The co-rank corresponds to the number of linearly independent $B$ to which the image of $\delta V_T$ is orthogonal.
In contrast to the case without the polarizability term found in \cite{Wu}, a differential equation is found by differentiation as $\frac{dE}{dt}$ remains in the resulting equation.
It is not yet clear what the full significance of this difference in form is beyond the observation that it represents that the singular controls, and thus potentially the set of landscape traps, possess a different structure in such cases meriting investigation.

A singular control may further be a singular critical point of the map $F$.
That is to say a singular control (satisfying eqn. (\ref{singular})) $w$ may have the property that:
\begin{align}
\la U_T^{\dagger} \delta U_T, U_T^{\dagger} \nabla J \big|_{U_T} \ra = 0, \ \ \forall \delta E
\end{align}
These are candidates for traps (i.e. local optima) in the landscape $F$ as they are controls for which $\nabla F\big|_{w} = 0$, i.e. critical points of $F$ for which $V_T[w]$ is a not critical points of $J$ (i.e. no a regular critical point).

\section{Singular Critical Points and Traps}

Not every singular control is a singular critical point of $F$ and not every singular critical point is a trap, i.e. a true local optima of $F$ in control space.
The analysis of which, if any, singular critical points are true traps requires an analysis of the Hessian index of the end-point map evaluated at singular critical points and appears prohibitively difficult by any  methods present in the quantum control literature.
With or without polarizability, which controls are singular does not depend on which function $J$ is being optimized but only on the underlying differential equation (Schr\"odinger equation (\ref{sch})) and the form of the Hamiltonian.
Insight can be gained by examining the derivative of $F$ by applying the chain rule:
\begin{align}
\frac{\delta F}{ \delta E} = \frac{d J}{d V_T[E]} \circ \frac{\delta{V_T}[E]}{\delta E}
\label{chain}
\end{align}
One sees that a control being singular can, but does not always, introduce a critical point of $F$; that is a control $E$ for which $\frac{\delta F}{\delta E} = 0$.
If a control $E$ is singular then $\frac{\delta{V_T}[E]}{\delta E}$ fails to be full rank.
However, only when $\delta U_T$ cannot vary (when all $\delta E$ are considered) in the direction of increasing $J$ (in direction of the gradient of $\nabla J$) is a critical point of $F$ introduced which would otherwise have been absent.
Such singular critical points are candidates for local optima on te landscape $F$.

\section{Numerical Evidence That Landscapes Including Polarizability Are Generically Trap-Free}

In order that numerical optimization could be tractably performed, we assumed a simplifying form of the control field $E$.
We assume that it is piecewise constant, but with many pieces so that general control fields can be very well approximated.
500 constant sections were used in all experiments,
This is in keeping with a well know theorem about approximating a general smooth function with a piecewise constant one \cite{nlc}.
Henceforth, the control will be represented by $\vec{E} \in \mathbb{R}^{500}$.
We used a simple randomized gradient ascent algorithm to probe the landscape.

The finial time propagator associated to the control $E$ without the polarizability term is:
\begin{align}
V_T: \vec{E} \mapsto U_T = \prod_{k=0}^{500} e^{(T/500)(iH_0+E_k iH_1)}
\end{align}
and with the polarizability term:
\begin{align}
V_T: \vec{W} \mapsto U_T = \prod_{k=0}^{500} e^{(T/500)(iH_0 + E_k iH_1 + E_k^2 iH_2)}
\end{align}
The algorithm used to optimize a given initial $E$ is essentially a randomize version of the now ubiquitous gradient ascent pulse engineering algorithm \cite{grape}.
The incarnation we modify is the standard gradient ascent method, rather than the conjugate gradient method often applied.

In order to assess the existence of traps in the landscape of a given Hamiltonian, the algorithm must be repeated many times with different initial random $\vec{E}$.
If the algorithm returns success regardless of the starting point, this indicates that the given Hamiltonian's landscape has no traps in the vicinity of attempted initial $\vec{E}$.

The algorithm (pseudo-code can be found in the online appendix materials) will always converge to a (potentially local) optima as the objective function $F$ is smooth and bounded and because only changes in $E$ which improve the fidelity are accepted in each iteration.
Furthermore, the randomization of the variation at each iteration of the algorithm precludes the possibility of getting stuck at a saddle (either at a singular or regular saddle point in control space) and is the reason for choosing this algorithm over the ordinary gradient algorithm.
This is because, after sufficiently many iterations, a $\delta E$ will be randomly generated which moves $E$ in a direction of increasing $F$ in control space, and thus any saddle will be escaped.
Any saddle has, by definition, such a direction, as opposed to a local optimum which does not.
The ordinary gradient method, if it lands exactly on a singular critical point, will cease to improve the objective as it has reached a point where the gradient is exactly zero despite.
The ordinary approach can also be very slow in the vicinity of saddle points as the gradient near these points is very small; even if $E$ does not exactly `land' on such a point exactly. It is highly improbable that an ordinary gradient search will land on a singular critical control due to inevitable tiny numerical errors during a search and their small volume in control space \cite{an1}.

\subsection{The Effect Of Adding A Polarizability Term To An Uncontrollable System with Traps}

It is known that the control landscapes of uncontrolable systems can contain traps \cite{rcb}.
Here we investigate if including polarizability can mitigate this effect.
A class of systems we study possess drift and control terms respectively given by:
\begin{align}
iH_0 & = J_x (i\sigma_x \otimes \sigma_x) + J_y (i\sigma_y \otimes \sigma_y) + J_z (i\sigma_z \otimes \sigma_z) \\ \nonumber
iH_1 & = i \mathbb{I} \otimes \sigma_z
\label{ranhei}
\end{align}
$H_1$ represents the effect of an external field in the z direction. This is a convenient choice of a parameterized family of systems for which traps exist.
The three assumptions are not all satisfied as these systems are not controllable, and thus it is not guaranteed that traps do no exist.
The failure of controllability can be confirmed by checking the standard Lie algebra rank condition.

\begin{figure}[H]
\centering
\caption{Typical Final Distribution of Fidelities Without Polarizability Indicating Trapping}
\includegraphics[width=0.3\textwidth]{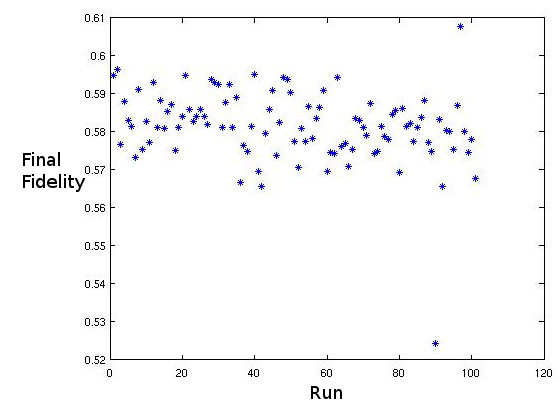}
\label{land1}
\end{figure}

1000 $(\vec{J}, G)$ tuples (with $||\vec{J}||=1$) were generated and the algorithm was run with 100 random initial conditions for each.
Almost all runs identified traps and converged to fidelity values less than $1$.
An example of the typical profile for the final value of fidelity when the algorithm terminated can be seen in figure (\ref{land1}).
In order to assess the effect of adding a polarizability term, 1000 further tuples $(\vec{J}, H_2,G)$ were generated and their landscapes were similarly analyzed.
It was found that, of 1000 initial $E$ for 1000 randomly generated $H_0$ and $H_2$ (with $H_0$ of the form \ref{ranhei}).

\setlength{\tabcolsep}{0.5em}
{\renewcommand{\arraystretch}{1.6}
\begin{table}[H]
\centering
\caption{Numerical Results}
\label{res}
\begin{tabular}{|l|c|c|}
\hline & \begin{tabular}[c]{@{}c@{}}Without\\ polarizability\end{tabular} & \begin{tabular}[c]{@{}c@{}}With\\ polarizability\end{tabular} \\ \hline
\begin{tabular}[c]{@{}l@{}} Fraction of pairs $(H_0, H_1)$ \\ which always converged \\ for all initial $E$
\end{tabular}
& 0\% & $99\%$ \\ \hline
\begin{tabular}[c]{@{}l@{}} For pairs $(H_0,H_1)$ with any \\ non-convergent controls: \\ the fraction of initial $E$ \\ which did not converge
\end{tabular}
& 99\% & 1.5 \% \\ \hline
\end{tabular}
\end{table}

We see that it is highly typical for the addition of a polarizability term to remove traps from the landscapes of systems which are uncontrollable through the dipole term alone.

\subsection{Typical Properties of Generic Systems with Polarizability}

In this section we assess if the addition of a polarizability term can introduce traps into the landscapes of systems known to have none in practice without a polarizability term.
We numerically analyze the landscape of more general systems which include polarizability.
1000 random tuples $(H_0,H_1,H_2,G)$ were generated and their landscapes similarly analyzed with 1000 runs each with random initial $E$.
These landscapes were found to be, as far as practical numerical optimization in concerned, trap free.
For all tuples generated, 1000 runs of optimization were completed with random initial $E$.
The initial $E$ were, as before, uniformly randomly generated to have $E(t) \in [-1,1]$ for all $t \in [0,T]$, but were unrestricted during the optimization.
All converged within practical time scales to fidelity above $0.95$ and no indication of trapping was observed.

100 runs starting from $E(t)=0$ were also completed with random tuples similarly generated as an attempt to test a claim analogous to that made in \cite{sscl} about the potential for the zero control to represent a trap.
Our systems are, however, qualitatively different from the one shown to possess a trap at the zero control, due to the polarizability term.
Identical convergence behavior was observed to the case with random initial control, which indicates that the zero control is not generically trapping for such systems.

\section{Size of $||E||$ During Gradient Ascent}

The term $H_2$ typically has size an order of magnitude less than $H_1$ in physical applications.
In situations where a control is exploiting the effect of $H_2$ to achieve the implementation of a desired gate $G$, one might reasonably conjecture that a small size polarizability $H_2$ entails a requirement for a very strong field or very high total field fluence.
The norm $E$ is plotted against iteration of algorithm (\ref{alg}) for three cases of 20 randomly generated triples $(H_0,H_1,H_2)$ in order to assess this possibility for impractical resource requirements for control via polarizability.
The three analyzed cases are those where $||iH_2||$ is approximately equal (the random generation function has the same expectation) to $||iH_0||$ (shown left), $||iH_2|| \sim \frac{1}{10} ||iH_2||$ (shown right) and $||iH_3|| \sim \frac{1}{100}$ (shown bottom).
\begin{figure}[]
\centering
\caption{$||E||$ against algorithm iteration in three cases: $||iH_2|| \sim ||iH_1||$, $\frac{1}{10}||iH_2|| \sim ||iH_1||$ and $\frac{1}{100}||iH_2|| \sim ||iH_1||$}
\includegraphics[width=\columnwidth]{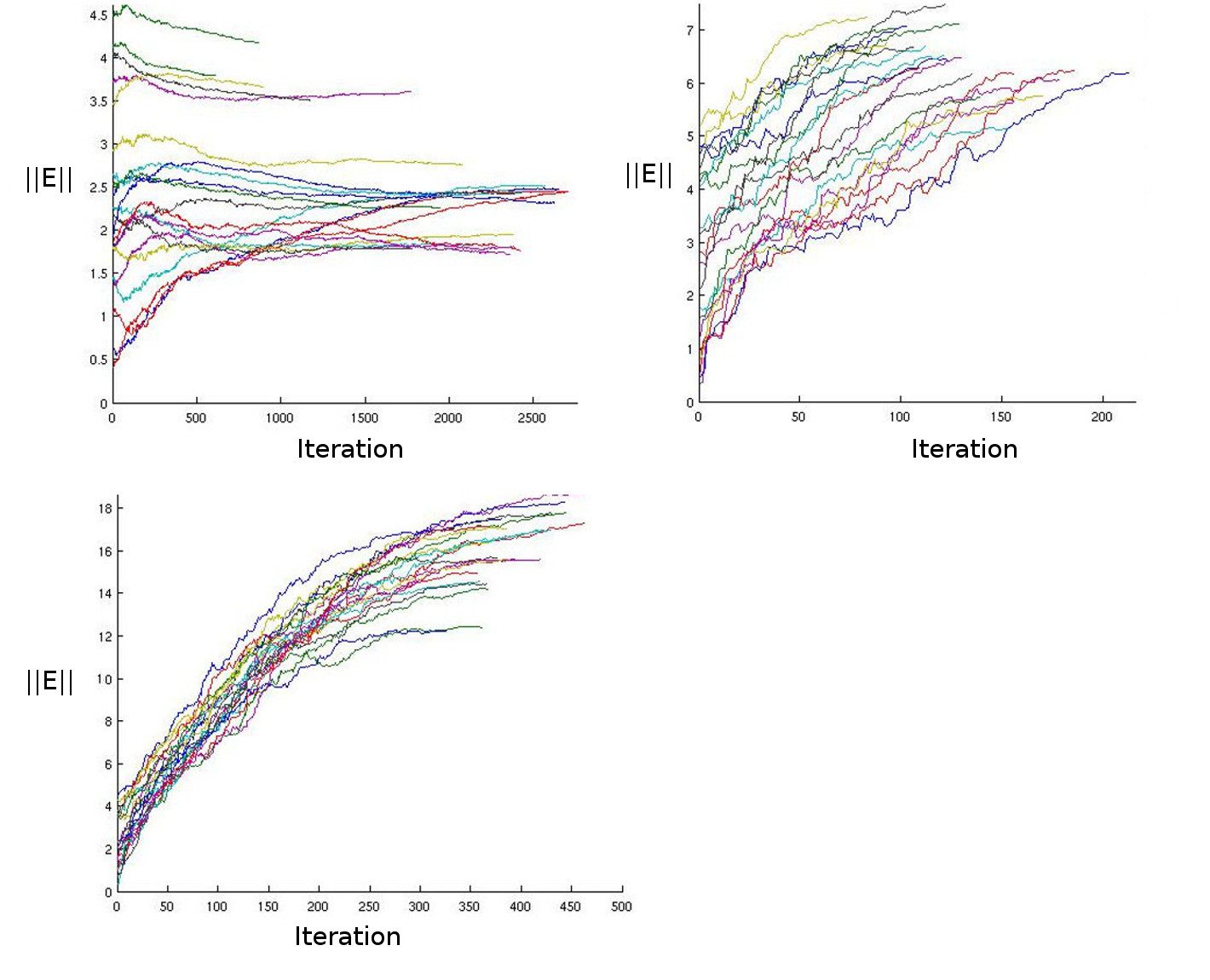}
\label{p1}
\end{figure}

One sees from the presented figures (\ref{p1}) that the final value of the total fluence does not depend sensitively on the size of the initial value of $E$.
One further sees that when $iH_1$ and $iH_2$ have similar sizes, the controls obtained seem not to increase in total fluence compared to the total fluence of the initial field.
This indicates that the effects of polarizability can be exploited for control without, at least for some conditions, requiring that the control field become large relative to the cases without the polarizability term.

\section{Numerical Assessment of The Neighborhood Of Singular Controls}

In this section we explore the neighborhood of singular controls to check for trapping behavior ni several types of system.
It is not known exactly what proportion of singular controls are singular critical controls, and what proportion of singular critical controls are traps.
Here we assess if singular controls play a significant role in determining the topology of critical points on quantum control landscapes for systems with polarizability.
Following the work in \cite{Wu, bon}, one can numerically solve a Schr\"odinger equation to obtain singular controls.
This can be achieved by substituting eqn. (\ref{sc}) into eqn. (\ref{polham}) and then substituting the resulting Hamiltonian into eqn. (\ref{sch}) to obtain the initial value problem:
\begin{align}
i\frac{d U_t}{dt} & = \left(H_0 -\frac{1}{2}\alpha(t) H_1 + \frac{1}{4}{\alpha(t)}^{2} H_2 \right) U_t \\
\nonumber
\alpha(t) & := \frac{\la U_t H_1 U_t^{\dagger}, B\ra}{\la U_t H_2 U_t^{\dagger}, B\ra}
\\
U_0 & = I
\label{singtraj}
\end{align}
the solutions to which are are singular trajectories emanating from the group identity at $t=0$.
One sees from eqn. (\ref{singtraj}) that the set of all singular trajectories through the identity is parameterized by $B \in \mathfrak{su}(n)$.
I.e. there is a single trajectory for each $B$ and a single $B$ for each trajectory.
From a numerical solution to eqn. (\ref{singtraj}), the corresponding singular control can be numerically obtained by substituting this numerically solution for a singular trajectory $U_t$ into eqn. (\ref{sc}).

In order to test if any given singular control $E$ is a trap, it is possible to explore the neighborhood of $E$ by evaluating $F[E +\delta E]$ for many small $\delta w$ and assessing the sign of $\delta F = F[E +\delta E] - F[E]$.
If two linearly independent $\delta E$ can be found such that $\delta F$ has different signs (i.e. one positive and one negative) then the point $E$ must be a saddle in control space rather than a trap.
Two types of system were assessed in this respect.

\subsection{Random $H_0,H_1,H_2$}

The first class of Hamiltonians assessed were those for which $H_0,H_1,H_2$ were chosen uniformly at random.
A typical singular control generated for this class of system can be seen below (\ref{singsf}).
This figure indicates that set of singular controls includes physically plausible electromagnetic fields which could be generated in laboratory practice.
This control was not critical.
However, upon examination, the singular critical controls exhibited no clear visually differentiating features when compared to the singular, non critical controls.
As such no example is shown.

10000 random tuples $(H_0,H_1,H_2,B,G)$ were generated and the corresponding singular control was found numerically by solving eqn. (\ref{singtraj}) in order to obtain a singular trajectory and thus, in turn generate a singular control.

In all cases generated, no traps were identified.
An average of 3.23 variations of the control were required to identify two which resulted in fidelity variations of opposite sign. 
Furthermore, the highest number of trial variations required for any singular control was 70.
This indicates that all singular controls examined are, at worst, saddles on the control landscape and that is easy to numerically confirm.

\begin{figure}[]
\centering
\caption{A Typical Singular Control As a Function Of Time}
\includegraphics[scale=0.25]{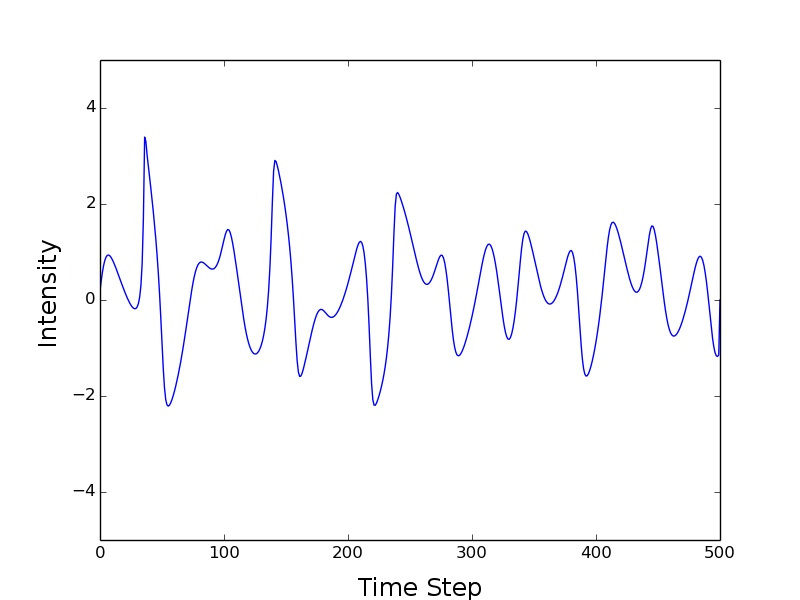}
\label{singsf}
\end{figure}

\subsection{Random $H_2$ and random coupling}

The second class of Hamiltonians assessed were those, as in eqn. (\ref{ranhei}), for which $\vec{J}$ and $H_2$ were chosen uniformly at random with $|J|=1$.
10000 random tuples $(\vec{J},H_2,B,G)$ were generated and the corresponding singular control was found numerically by solving eqn. (\ref{singtraj}) in order to obtain a singular trajectory and thus a singular control, as before.
In all cases generated, no traps were found.
The highest number of trial variations required for any singular control was 200.
This indicates that all examined singular controls examined are, at worst, saddles on the control landscape and that this fact is also easy to numerically confirm.

Upon visual examination of many singular controls, they seem to exhibit characteristic features.
Most notably, there are two distinct classes.
The first class are all physically plausible as fields which could be created in the lab or which could be encountered during a gradient ascent simulation as they are smooth and bounded.
Controls in the second class all posses at least one blow-up point where the control becomes both unbounded and discontinuous (an infinite jump discontinuity similar to the reciprocal function $f(x)=\frac{1}{x}$ at $x=0$).
As such, they are clearly excluded from physical consideration.

\section{Numerical Assessment of The Neighborhood Of Singular Critical Controls}

For each element $B\in\mathfrak{su}(4)$, there is a singular control defined by eqn. (\ref{singtraj}).
Using randomized gradient decent (similar to Fig (\ref{alg}), and thus omitted) for $B$, one can produce singular critical controls, rather than just singular ones.
We then optimize over $B$ to achieve $U_T^{\dagger} \nabla J\big|_{U_T}$ co-linear (to a numerical tolerance of $0.001$ radians) to $B$.
In the case of $\mathfrak{su}(4)$, this is a minimization over the 15 parameters of $\mathfrak{su}(4)$.
One parameter can be discarded as the formula for a singular control doesn't depend on the norm of $B$ due to the linearity of the numerator and denominator.
As such it is possible to restrict the search to unit norm $B$.
We found that this search did not always succeed in finding a singular critical control; only about 5\% of searches succeeded. This suggests at least of one the following: not all systems have any singular critical controls, that the set of singular critical controls is very small in the set of singular controls or all routes to the set of singular critical controls in the space of all $B$ is blocked by many local optima.

We studied the structure of the control landscape in systems with randomly generated $(H_0,H_1,H_2,B,G)$.
In order to analyze the possibility for a singular critical control $E$ to be a trap on the control landscape, many small variations $E^{'}:=E+\delta E$ were generated.
For each $E'$ a randomized gradient ascent was initiated.
If a singular critical $E$ were a true trap, rather than a saddle, this would be identified by at least some gradient ascent runs returning back to $E$ (or a control near to $E$ of the same fidelity) when initiated from $E + \delta E$.
If the fidelity during the run reached one, then $E$ cannot be a trap.
100 tuples $(H_0,H_1,H_2,B,G)$ were generated and for each 100 singular critical controls were generated.
For each singular critical control $E$, 200 points $E^{'}$ in the neighborhood of $E$ were generated and for each a gradient ascent run was completed.
The norm of $\delta E$ was chosen to be random within $[0, 0.001]$ so as to ensure exploring closer to the candidate trap and the behavior around it.
We found that all runs converged to $J[U_T]=1.0$ and displayed similar convergence rates as those seen when initial controls were chosen at random from the whole control space.

\section{The Effect Of A Polarizability Term On A Known Second Order Trap}

In this section we assess the potential for the addition of a polarizability term to removed traps from landscapes know to possess some.
One example of a second order, i.e. negative definite Hessian, is known for the control of the propagator.
This example is a three level system.
This example can be found in \cite{sscl} and in \cite{riv} where it is referred to as: `system E'.
This system is defined as:
\begin{align}
H_0 & =
\begin{bmatrix}
1+\alpha & 0 & 0 \\
0 & 1 & 0 \\
0 & 0 & 2
\end{bmatrix}, \ \
H_1 =
\begin{bmatrix}
-a & -1 & 0 \\
-1 & -b & -1 \\
0 & -1 & -c
\end{bmatrix} \\ \nonumber
W & =
\begin{bmatrix}
e^{i\theta} & 0 & 0 \\
0 & -ie^{-i\phi} & 0 \\
0 & 0 & -ie^{i\phi}
\end{bmatrix}
\exp\left(-iH_0 T/\hbar \right)
\end{align}
where:
\begin{align}
a = 5\sqrt{\frac{2}{3}}, \ \
b = 4, \ \
c = 1, \ \
\theta = \frac{2\pi}{3}, \\ \nonumber
\phi = \frac{-3\pi}{4}, \ \
T= \frac{\pi}{\alpha} = 1000, \ \
\alpha = \frac{\pi}{1000}
\end{align}
as in \cite{riv}.

We generated 1000 random polarizability matrices $H_2 \in \mathfrak{su}(3)$ ($\frac{1}{10}$ of the norm of $H_1$), for each a random initial field was generated in the vicinity of the zero field.
The initial fields were generated to decrease in size with each overall iteration, converging to the zero field as more and more polarizability matrices were tested.
It was found that, for initial pulses arbitrarily close to the zero field, all gradient ascent runs converged to $1$.
As such, we can conclude that the trapping effect observed in \cite{riv} has been counteracted by the addition of a polarizability term, and that this counteraction happens similarly for almost all, if not all, values of the polarizability term.

\section{Conclusions, Outlook And Further Work}

We have shown that including a polarizability term in the Hamiltonian, and thus moving beyond the standard dipole approximation by including additional physically relevant terms in the Hamiltonian, a change in the character of quantum control landscapes is seen in some physically relevant three and four level cases.
This effect has been numerically confirmed to be highly typical and not to depend on any of the details of the choice of polarizability matrix $iH_2$ for many pairs $(H_0,H_1)$.
We have also shown that there is a clear theoretical difference between the formula for the singular controls in the cases with and without the polarizability term.

There are three central conclusions:
\begin{enumerate}
\item Including a polarizability does not introduce traps in to the landscape for typical triples $(iH_0, iH_1, iH_2)$.
\item Including a random polarizability term can remove traps from the landscapes for a class of an otherwise uncontrollable systems.
\item Including a random polarizability matrix can remove existing traps from the landscape of a specific three level system known to possess a trap at the zero field.
\end{enumerate}

It appears that neither with nor without $H_2$, singular controls generically (over all Hamiltonians $(iH_0, i_H1)$) introduce traps in quantum control landscapes in except a few very specific mathematical scenarios.
However, a general proof of this is \emph{still} illusive as is an understanding of the most general class of systems with trap free control landscapes.

This work serves to bolster the claim that trap free landscapes are ubiquitous in the practice of quantum control.
The addition of novel physically meaningful terms, such as the polarizability, which in reality is always present (but may be negligibly small), serves only to further improve the trap free status of the landscapes of such systems.
It was also shown that adding polarizability can render a quantum control landscape trap free, which was not trap free with out this term via the restoration of controllability through the polarizability term $H_2$ in a system in which $H_1$ alone was not sufficient.

In this work, an algorithm was devised to search for singular controls and analyses if they are traps or not by examining $F$ in a neighborhood of any singular controls found.
Effectively, this process is estimating the eigenvalues of the Hessian of the end-point map at and near to singular critical points.
In \cite{Wu} no trapping singular controls were found in the case of linear coupling and the control of the density matrix.
Our work extends this work in two directions.
Firstly, we study the control of the full quantum propagator $U_t$.
Secondly, we study the role of non-linear coupling to an electromagnetic field.
Future work will include repeating the numerical analysis of \cite{Wu} in the case of the control of the density matrix  and the observable maximization task for systems with a polarizability term and even higher order coupling terms.

For control fields with frequencies above about the x-ray range, coupling terms to the spacial derivatives of the control field become physically relevant. This is because the wavelength of the control field can become comparable to the spacial extent of the electronic wave function, thus the field varies significantly within the boundary of the system being controlled. While such fields and coupling mechanics are not currently being exploited in laboratory quantum control, lasers capable of producing fields within the relevant frequency band are being developed \cite{xray}. The effect of such additional coupling terms, which would compliment the terms in the polarizability terms, is as of yet unstudied in the control literature.
The effect of such additional coupling terms on control landscapes is not known.

Finally, we conjecture that including more terms in the expansion (\ref{expa}) will have the general effect of removing traps from the landscape by adding novel mechanisms of control via coupling to an external field.
This will form the basis for further analytical and numerical investigation.

\section{Acknowledgments}

Thanks to Liang-Yan Hsu for several helpful comments on the physics of molecular polarizability.
Thanks to NSF (grant no. CHE-1464569 for funding Benjamin Russell, to the Army Research Office (Grant No. W911NF-13-1-0237) for funding Herschel Rabitz.

\section*{Bibliography}
\bibliographystyle{unsrt}
\bibliography{main}

\appendix*
\section{Randomized Gradient Algorithm}

\begin{figure}[H]
\centering
\begin{algorithmic}
\caption{Randomized Gradient Algorithm}
\label{alg}
\State $ \vec{E} \gets \text{random\_vec}(500)$
\State $U_T \gets V_T[\vec{E}]$
\State fidelity $\gets \frac{1}{16}\left|Tr(G^{\dagger}U_T) \right|^2$
\\
\While{fidelity $\leq 0.95$}
\While{(trial\_fidelity $\leq$ fidelity)}
	\State $\Delta \vec{E} \gets \epsilon \times \text{random\_vec}(500)$
	\State $\vec{E}_{\text{trial}} \gets \vec{E} + \Delta \vec{E}$

    \State $U_T \gets V_T[\vec{E}_{\text{trial}}]$
	\State trial\_fidelity $\gets \frac{1}{16}\left|Tr(G^{\dagger}U_T) \right|^2$
\\
	\If{tries $\geq 1000$} \Return Failure \EndIf
\\
    \State tries $\gets$ tries $+1$
\EndWhile
\\
\State tries $\gets 0$
\State fidelity $\gets trial\_fidelity$
\State $\vec{E} \gets \vec{E} + \Delta \vec{E}$
\EndWhile
\\ \\
\Return Success
\end{algorithmic}
\end{figure}

\end{document}